\documentclass[a4paper,twocolumn,showpacs,amsmath,amssymb,prl]{revtex4}
\usepackage{graphicx}
\usepackage{dcolumn}

\begin{document}

\title{Experimental observation of fractal modes in unstable optical resonators}

\author{J.A. Loaiza}
\email{javier@molphys.leidenuniv.nl@}
\author{E. R. Eliel}
\author{J. P. Woerdman}
\affiliation{Huygens Laboratorium, Universiteit Leiden, P.O. Box 9504, \\
2300 RA Leiden, The Netherlands}
\homepage{http://www.molphys.leidenuniv.nl/qo/index.html}

\begin{abstract}
We use a spatially resolved  cavity ring-down technique to show
that the 2D eigenmode of an unstable optical cavity has a fractal
pattern, i.e. it looks the same at different length scales. In
agreement with theory, we find that this pattern has the maximum
conceivable roughness, i.e., its fractal dimension is 3.01 $\pm$
0.04. This insight in the nature of unstable cavity eigenmodes may
lead to better understanding of wave dynamics in open systems, for
both light and matter waves.
\end{abstract}
 \maketitle

Conventional lasers are based upon stable optical cavities in
which the light rays that bounce between the two mirrors are
trapped forever and leakage around the mirror edges is negligible.
In wave-optics language this picture of bouncing rays corresponds
to the well known Hermite-Gaussian eigenmodes\cite{BookSiegman}.
In contrast, in an unstable cavity (UC),  consisting for instance
of two convex mirrors that face each other, the combination of the
curvatures of the mirrors and their separation is such that the
bouncing rays run away from the axis towards the mirror edges and
ultimately escape;  the escape rate is determined by the
magnification $M$ of the system.  The associated eigenmodes fill
the entire volume of the cavity; this has the advantage of optimum
energy extraction if an active medium is placed inside the cavity
to realize a laser. Another advantage of an unstable-cavity laser
is that discrimination against higher-order transverse  modes is
much better than for a stable-cavity laser; perfect single-mode
oscillation is very easily achieved\cite{BookSiegman}. These
advantages have been known since the early days of laser
physics\cite{ArtSiegman1}, whereas other surprising properties of
the UC laser have emerged more recently. In particular we refer
here to the much debated phenomenon of UC excess quantum
noise\cite{cheng,vanEijk,new1,vanderlee,petterman} and the
prediction that their transverse eigenmodes are
fractals\cite{karman1,karman2}, i.e., they are invariant under
magnification\cite{mandelbrot}. While the former effect has been
experimentally demonstrated, the latter is still being pursued.
What we report here is the first experimental observation of the
fractal nature of UC eigenmodes in the optical regime; these
results may stimulate extension to matter waves.

On an intuitive level the origin of  the fractality is rooted in
two arguments: (i) an unstable cavity has a round-trip
magnification $M>1$. (ii) The cavity eigenmode must, by
definition, be invariant upon round-trip  propagation. The
combination of arguments (i) and (ii) implies that an UC eigenmode
must be self-similar (i.e. fractal). This line of reasoning must,
however, be handled with care since the concept of magnification
has its origin in ray-optics, whereas an eigenmode is a
wave-optics concept\cite{courtial}; moreover, these arguments do
not lead to predictions for the fractal dimension.

The starting point to describe the mode structure of an UC is the
so-called Virtual Source (VS) method, originally developed by
Horwitz\cite{horwitz} and Southwell\cite{southwell} . In the VS
method a two-mirror UC is unfolded to create a corridor of virtual
images of the confining aperture (e.g. the smallest of the two
mirrors). A plane wave is injected at the far end of the corridor
and the eigenmode is formed by superposition of the diffracted
patterns produced by the sequence of virtual apertures. This
superposition (i.e. the eigenmode) contains thus rich spatial
structure that depends on the shape of the confining aperture.

Very recently,  Berry and co-workers formulated an
\textit{analytical} theory for fractal
eigenmodes\cite{berry1,berry_vansaarloos} and excess quantum
noise\cite{berry2}, by introducing further approximations to the
VS method. This is an important development since it carries the
notion of a fractal pattern beyond that of a phenomenological
description, yielding insight and allowing predictions; fractality
then presents a convenient test of the theory as a whole.
Specifically, the theory predicts that in the asymptotic limit
$N_F \rightarrow \infty$, the fractal dimension $D$ of the
fundamental eigenmode is 3 for a wide variety of polygon-shaped
confining apertures\cite{berry1} (e.g. a triangle). Here $N_F
\equiv a^2 /\lambda L$ is the Fresnel number, where $a$ is the
(linear) size of the aperture, $L$ is the length of the cavity and
$\lambda$ is the wavelength of light. This prediction is very
surprising since it implies that the 2D transverse intensity
profile, when viewed as a mountain landscape, has the maximum
topologically allowed roughness. Contrary to what might be
expected, diffraction does \textit{not} smooth out this profile;
in fact, diffraction is the \textit{cause} of the phenomenon.

We stress that experimental verification of this prediction is
essential, since the validity of the theory may be questioned at
various levels. First, all theoretical efforts so far, be it
numerical or analytical, are based upon standard diffraction
theory; this is an inherently approximate theory that derives its
simplicity (as compared to solving Maxwell's equations) from using
inconsistent boundary conditions\cite{stamnes}. Although this is
hardly ever a problem in practical cases, questions remain when
dealing with an unusual problem such as the UC. On top of this,
the  VS method is a not a straightforward implementation of
diffraction theory but a paraxial model that combines some
elements of diffraction theory in an \textit{ad hoc} fashion.
Finally, the adaptation of the VS method by Berry \textit{et al.}
uses additional assumptions in order to arrive at an analytical
result: the asymptotic limit $N_F\rightarrow \infty$ must apply
and multiple edge reflections as well as corner-point diffraction
are neglected.

In the experiment we study the lowest-loss eigenmode of an
\textit{empty} unstable cavity in order to avoid the complications
associated with a gain medium. As we do not know how to
selectively excite this fundamental mode, we let the cavity do the
selection process. Briefly, we use a variant of the
cavity-ring-down technique originally developed for molecular
spectroscopy\cite{cringdown, bretenaker}. We inject laser light
with an arbitrary transverse profile into the cavity (we do not
need to mode match the input beam to our cavity), rapidly
switch-off the laser, and then record the spatial profile of the
light emitted by the cavity after a certain ring-down
time\footnote{Of dominant concern is the light emitted by the
diode laser after it has been nominally turned off ('afterglow'),
for instance due to the recombination of remaining charge carriers
in the p-n junction. This is important because the intracavity
intensity of the UC has been reduced by many orders of magnitude
by the time that the pattern stabilizes. Obviously, one requires
any light unintentionally injected in the cavity at that time to
be even weaker. As a result of careful optimization of the
experimental parameters we obtain on/off ratios of the injected
light up to $10^8$\label{fn:afterglow}}.

Because we  work with an unstable cavity, the round-trip losses
are high, in our case of the order of a factor of 2 per
round-trip. This is in stark contrast with the usual situation
encountered in cavity ring-down spectroscopy where the round-trip
losses are exceedingly small. Consequently, the time scale of our
experiment is of the order of nanoseconds, instead of
microseconds. Since we are specifically interested in details of
the spatial distribution of the light transpiring through the
cavity mirrors,  we  use a camera instead of an integrating
detector.

\begin{figure}
\includegraphics{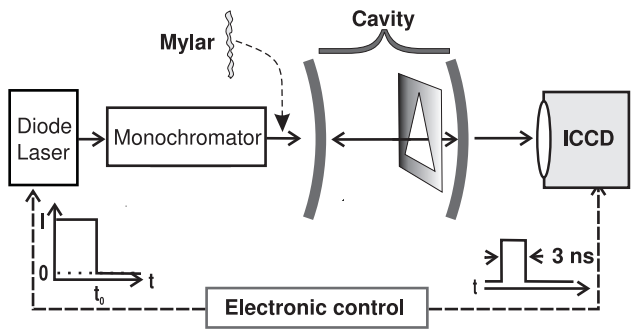}
\caption{\label{fig:fig1}Schematic diagram of our experiment.
Light from a diode laser is filtered by a monochromator and
excites the cavity. We use a convex-concave unstable cavity of
length $L=50$ cm and round-trip magnification $M=1.68$. The
metallic polygonal aperture is located near the exit mirror, and
is mounted on a 2-axis translation stage for purposes of
alignment. A Mylar film can be placed in the beam path, in order
create a new input distribution for the cavity.}
\end{figure}

 We use a positive branch unstable cavity (Fig. \ref{fig:fig1})
consisting of a front convex mirror\footnote{For the convex mirror
we use a concave mirror (radius of curvature 6) on a glass
substrate ($n=1.5$) in a reversed way.\label{fn:mirror}}
(effective radius of curvature $R_2=-4$ m), and a back concave
mirror ($R_1=10$ m), separated by a distance of 50 cm. Both
mirrors have a reflectivity coefficient of 0.9. The confining
aperture is positioned near the back mirror; it consists of a
regular-polygon shaped hole in a metal sheet. Here we report
results for triangular and hexagonal apertures whose typical
Fresnel numbers ($N_F=a^2/ \lambda L$) range around 90, with $a$
taken as the radius of the polygon's inscribed circle, typically
around 6 mm. The oscillatory behaviour of the cavity round-trip
propagator is characterized by means of the parameter $A \equiv 2
\pi M F$\footnote{Note that $A=2 \pi N_{coll}$, where $N_{coll}$
is the so called collimated Fresnel number \cite{BookSiegman};
note also that Berry \textit{et al}.
\cite{berry1,berry_vansaarloos} give an expression for $A$
tailored to the case of a confocal UC.}, where $F\equiv N_F/2g_2$,
$g_2=1-L/R_2$, and $R_2$ is the radius of curvature of the front
mirror\cite{horwitz} . In the present case the parameter $A$ takes
on values around 500. The confining aperture is 1-1 imaged on the
photocatode of a single-photon sensitive camera, whose exposure
time ($\approx 2$ ns) is shorter than the round-trip time of the
cavity ($\approx 3$ ns).

Light from a 75 mW laser diode (Sharp LT024MD0) at $\lambda$=786.3
nm is injected into the cavity through one of its mirrors after
first being spectrally filtered by means of a monochromator having
a 0.3 nm wide transmission window (Fig.1). Due to the low finesse
of the UC we do not need to tune our laser to one of its
resonances.  The experiment consists of rapidly switching-off the
diode laser, and after a short delay, making an exposure by
triggering the photocathode and  intensifier of the single-photon
sensitive camera(Princeton Instruments Pi-Max). Many subsequent
exposures are superimposed on the camera until the signal-to-noise
ratio of the acquired image is sufficiently large, followed by
read-out of the camera. For the longest delay of our experiment
($\approx 30$ ns) we have to accumulate typically $10^5$ exposures
to arrive at a spatial pattern with sufficient detail.

\begin{figure}
\includegraphics{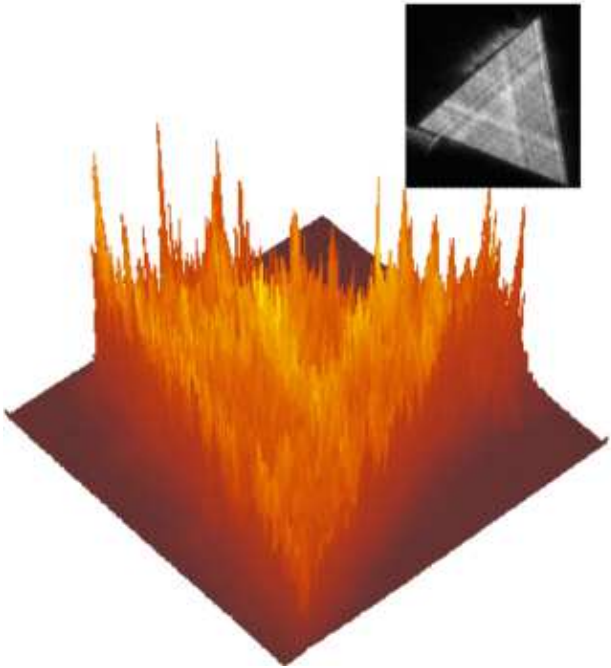}
\caption{\label{fig:fig2}3D landscape (bottom) and 2D (top)
snapshot of the measured intracavity intensity distribution of the
fundamental eigenmode of an unstable cavity ($M$=1.68, $N_F$=90),
taken 7 round-trips after the start of the decay. The camera has
512$\times$512 pixels, and a dynamical range of 16 bits.}
\end{figure}

Experimentally, we find that for our configuration (Fig.
\ref{fig:fig1}), the shape of the intra-cavity intensity
distribution stabilizes after approximately 7 round-trips. A
typical result of a 2D intensity image, together with its 3D
representation, is shown in Fig. \ref{fig:fig2}. The 2D picture
shows, with fascinating detail, the high-degree of symmetry and
structural self-similarity of the intensity pattern, while the
corresponding 3D landscape emphasizes its very abrupt topography.
It is worth mentioning that Mandelbrot\cite{mandelbrot} associates
fractal dimensions up to 2.4 to earthy looking landscapes;
therefore we expect a higher fractal dimension  for the weird
landscape shown in Fig. \ref{fig:fig2}.

We  use  the Fourier-transform method to evaluate the fractal
dimension of the intensity profile since this allows a direct
comparison with the analytical
theory\cite{berry1,berry_vansaarloos}. The method is based on the
fact that the power spectrum of a fractal function $f(x)$ follows
a power law $P(k) \approx k^{-b}$, and that the phase of its
Fourier components is random; the fractal dimension $D$ can be
obtained from the exponent $b$ by using the
relation\cite{falconer,berry_lewis} $2D=5-b$. It does not matter
whether we take for $f(x)$ the optical intensity or the field
amplitude, both choices lead to the same value of
$D$\cite{berry_fractal_boxes}.

\begin{figure}
\includegraphics{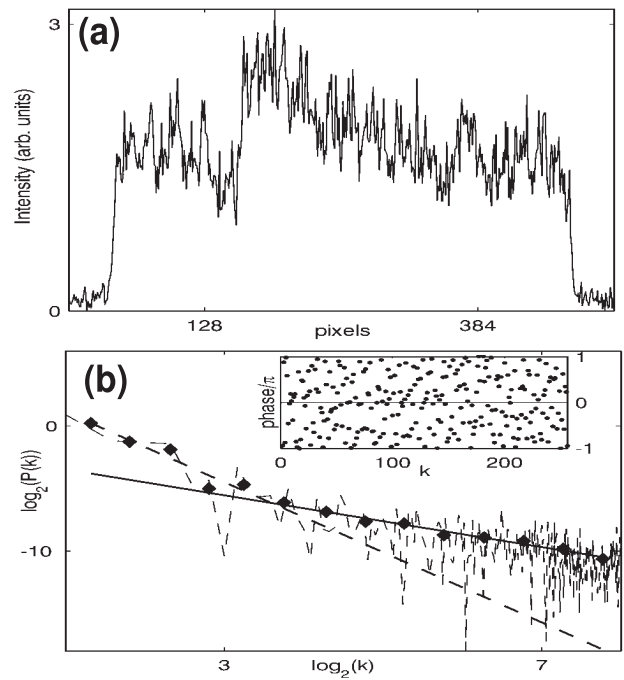}
\caption{\label{fig:fig3}Analysis of a fractal intensity pattern
taken 9 roundtrips after the start of the decay. Frame (a) shows a
typical 1D cut through the triangular pattern, while frame (b)
shows the corresponding normalized spatial power spectrum. The
thin dashed line shows the power coefficients of frame (a), the
diamonds indicate the average power over bands of logarithmically
constant length, equal to $\sqrt{2}$. The straight dashed line
represents a fit to the low-frequency data,  while the solid line
shows a fit to the high-frequency data; the latter fit corresponds
to the fundamental mode and yields a slope $b$=0.98, i.e. a
fractal dimension $D_{cut}$=2.01. The inset shows the phase
(normalised to $\pi$) of the Fourier transform components.}
\end{figure}

We follow this procedure for 1D cuts of a typical asymptotical
intensity profile. Fig. \ref{fig:fig3}a shows a 1D cut taken 9
roundtrips time after laser shut-down, while Fig. \ref{fig:fig3}b
shows its spatial power spectrum. This spectrum shows many
"spikes" due to the discrete nature of the virtual sources; these
have been smoothed by averaging over bands of logarithmically
constant length\cite{yates_and_new}. Apparently the power spectrum
is described by two power-law contributions dominating at low and
high spatial frequencies, respectively. Yates and
New\cite{yates_and_new} have shown that the higher the order of a
UC eigenmode, the steeper the power law of its spatial power
spectrum is. This leads us to associate the low spatial-frequency
range in Fig. \ref{fig:fig3}b with residual light in higher-order
modes. \textit{The high-frequency contribution is then associated
with the fundamental eigenmode}; a fit of the latter contribution
yields $D_{cut}=2.01\pm0.04$ for the fundamental mode.  The
fractal dimension $D$ of the 2D landscape follows directly from
this value, since $D=D_{cut}+1$, yielding $D\approx 3$, in
excellent agreement with theory\cite{berry1}. This value of the
fractal dimension implies that the intensity landscape has the
maximum allowed roughness. The inset in Fig. \ref{fig:fig3}b shows
that the Fourier components of the 1D cut have indeed random
phases, as required for a fractal curve.

\begin{figure}
\includegraphics{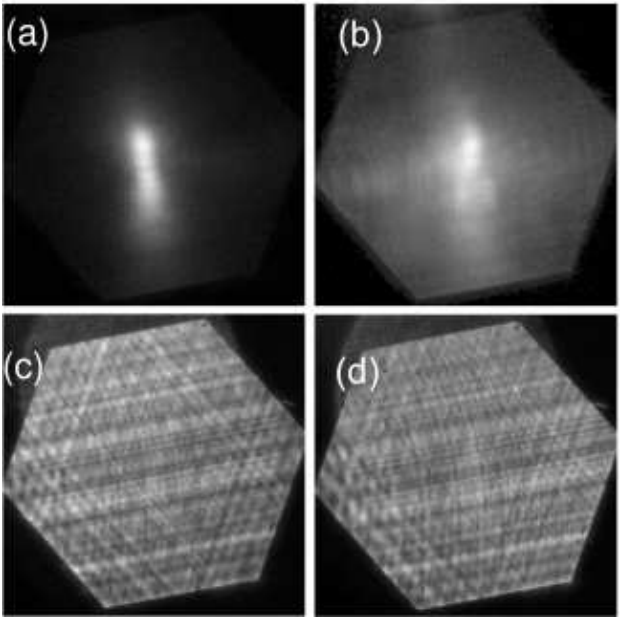}
\caption{\label{fig:fig4}Frames (a) and (b) show  the initial
distribution of light injected into the cavity without and with a
thin sheet of Mylar as a diffuser, respectively; frames (c) and
(d) show the corresponding pictures taken after 7 round trips.}
\end{figure}

 To ascertain that our cavity ring-down method indeed
selects a single cavity mode,  we have investigated whether the
observed asymptotical spatial pattern is sensitive to changes in
the initial conditions, i.e., to the spatial pattern of the
injected light. To do so, we have introduced a piece of Mylar film
as diffuser between the monochromator and the cavity, thereby
creating a totally different distribution of input phase and
amplitude.  Figure \ref{fig:fig4} shows, for a hexagonal aperture
inside the UC, the initial and asymptotic profiles of the
intensity distribution using a narrow injection beam, and a beam
that has been diffused by the thin Mylar film before entering the
cavity. The asymptotical spatial patterns are very similar,
implying that they correspond to a single mode.

Our experimental validation of the analytical theory of eigenmodes
of UC's\cite{berry1, berry_vansaarloos, berry2} allows its use in
a broader context. This is important since UC lasers play an
important role in laser physics; in particular microlasers. The
reason is that microlasers (e.g. semiconductor lasers) must be
efficient, which implies that the gain must be localized, e.g., to
a $\lambda$-sized transverse dimension; pumping outside this
region is a waste. Because of the gain localization there is gain
guiding in addition to the index guiding that is used to realize
the $\lambda$-sized optical confinement in the first place. The
(unavoidable) combination of gain and index guiding leads to a
cavity that is effectively slightly unstable, i.e., transversely
open to the outside world; and this increases the quantum noise of
the device \cite{vanderlee}. A better insight in the nature of UC
eigenmodes may lead to device architectures that minimize this
excess noise.

Finally, the issues discussed in this Letter can also be extended
to UC's (or open systems in general) for matter waves. As
demonstrated recently, spherical mirrors for atoms can be
made\cite{milner, friedman} so that an atom-optics based UC is
within reach. In fact, an UC for electron waves has already been
demonstrated, as a mesoscopic device based upon a 2D electron gas;
the unstable nature of this UC determines the conductance of a
quantum point contact placed inside\cite{heller,katine}. It
remains to be seen what the consequences of "fractal matter waves"
are for the operation of such devices.

\begin{acknowledgments}
This work is part of the research program of the 'Stichting voor
Fundamenteel Onderzoek der Materie' (FOM).
\end{acknowledgments}

\end{document}